\documentclass[letterpaper]{article}
\usepackage{aaai}
\usepackage{times}
\usepackage{helvet}
\usepackage{courier}
\usepackage{graphicx}
\usepackage{xcolor}
\begin{document}
%
\title{\#MeToo on Campus: Studying College Sexual Assault at Scale \\ using Data Reported on Social Media}
\author{Viet Duong, Phu Pham, Ritwik Bose, Jiebo Luo\\
Department of Computer Science\\
University of Rochester\\
\{vduong, ppham2\}@u.rochester.edu\\
\{rbose, jluo\}@cs.rochester.edu\\
}
\maketitle
\begin{abstract}
\begin{quote}
Recently, the emergence of the \#MeToo  trend  on  social media has empowered thousands of people to share their own sexual harassment experiences. This viral trend, in conjunction with the massive personal information and content available on Twitter, presents a promising opportunity to extract data-driven insights to complement the ongoing survey-based studies about sexual harassment in college. In this paper, we analyze the influence of the \#MeToo trend on a pool of college followers. The results show that the majority of topics embedded in those \#MeToo tweets detail sexual harassment stories, and there exists a significant correlation between the prevalence of this trend and official reports on several major geographical regions. Furthermore, we discover the outstanding sentiments of the \#MeToo tweets using deep semantic meaning representations and their implications on the affected users experiencing different types of sexual harassment. We hope this study can raise further awareness regarding sexual misconduct in academia.
\end{quote}
\end{abstract}

\section{Introduction}

Sexual harassment is defined as "bullying or coercion of a sexual nature, or the
unwelcome or inappropriate promise of rewards in exchange for sexual favors."~\footnote{www.en.wikipedia.org/wiki/Sexual\_Harassement} 
In fact, it is an ongoing problem in the U.S., especially within the higher education community. According to the  National Sexual Violence Resource Center (NSRVC), one in five women and one in sixteen men are sexually assaulted while they are attending college.~\footnote{https://goo.gl/rgvYH2} 
In addition to the prevalence of campus sexual harassment, it has been shown to have detrimental effects on student's well-being, including health-related disorders and psychological distress \cite{JTS:JTS213,doi:10.1177/0146167205284281}. However, these studies on college sexual misconduct  
usually collect data based on questionnaires from a small sample of the college population, which might not be sufficiently substantial to capture the big picture of sexual harassment risk of the entire student body.

Alternatively, social media opens up new opportunities to gather a larger and more comprehensive amount of data and mitigate the risk of false or inaccurate narratives from the studied subjects. On October 15 of 2017, prominent Hollywood actress Alyssa Milano, by accusing Oscar-winning film producer, Harvey Weinstein, for multiple sexual impropriety attempts on herself and many other women in the film industry, ignited the "MeToo" trend on social media that called for women and men to share their own sexual harassment experience. According to CNN, over 1.7 million users had used the hash-tag in 85 countries.~\footnote{http://www.cnn.com/2017/10/30/health/metoo-legacy/index.html} Benefiting from the tremendous amount of data supplied by this trend and the existing state-of-the-art semantic parser and generative statistical models, we propose a new approach to characterizing sexual harassment by mining the tweets from college users with the hash-tag \#metoo on Twitter.

Our main contributions are several folds. We investigate campus sexual harassment using a big-data approach by collecting data from Twitter. We employ traditional topic modeling and linear regression methods on a new dataset to highlight patterns of the ongoing troubling social behaviors at both institutional and individual levels.
We propose a novel approach to combining domain-general deep semantic parsing and sentiment analysis to dissect personal narratives.

\section{Related Work}
Previous works for sexual misconduct in academia and workplace dated back to last few decades, when researchers studied the existence, as well as psychometric and demographic insights regarding this social issue, based on survey and official data~\cite{FITZGERALD1988152,doi:10.1207/s15324834basp17042,doi:10.1111/j.1540-4560.1982.tb01912.x}. However, these methods of gathering data are limited in scale and might be influenced by the psychological and cognitive tendencies of respondents not to provide faithful answers~\cite{doi:10.1177/0149206312455245}.

The ubiquity of social media has motivated various research on widely-debated social topics such as gang violence, hate code, or presidential election using Twitter data \cite{blandfort2019multimodal,elsherief2018hate,DBLP:journals/corr/MaguJL17,DBLP:journals/corr/WangFZL16}. Recently, researchers have taken the earliest steps to understand sexual harassment using textual data on Twitter. Using machine learning techniques, Modrek and Chakalov (2019) built predictive models for the identification and categorization of lexical items pertaining to sexual abuse, while analysis on semantic contents remains untouched \cite{modrek2019metoo}. Despite the absence of Twitter data, Field et al. (2019) did a study more related  to ours as they approach to the subject geared more towards  linguistics tasks such as event, entity and sentiment analysis \cite{field2019contextual}. Their work on event-entity extraction and contextual sentiment analysis has provided many useful insights, which enable us to tap into the potential of our Twitter dataset.  

There are several novelties in our approach to the \#MeToo problem. Our target population is restricted to college followers on Twitter, with the goal to explore people's sentiment towards the sexual harassment they experienced and its implication on the society's awareness and perception of the issue. Moreover, the focus on the sexual harassment reality in colleges calls for an analysis on the metadata of this demographics to reveal meaningful knowledge of their distinctive characteristics \cite{DBLP:journals/corr/HeML16}.

\section{Dataset}
\subsection{Data Collection}
In this study, we limit the sample size to the followers identified as English speakers in the U.S. News Top 200 National Universities. We utilize the Jefferson-Henrique\footnote{https://git.io/JvUnQ} script, a web scraper designed for Twitter to retrieve a total of over 300,000 \#MeToo tweets from October 15\textsuperscript{th}, when Alyssa Milano posted the inceptive \#MeToo tweet, to November 15\textsuperscript{th} of 2017 to cover a period of a month when the trend was on the rise and attracting mass concerns. Since the lists of the followers of the studied colleges might overlap and many Twitter users tend to reiterate other's tweets, simply putting all the data collected together could create a major redundancy problem. We extract unique users and tweets from the combined result set to generate a dataset of about \textit{60,000} unique tweets, pertaining to \textit{51,104} unique users. 

\subsection{Text Preprocessing} 
We pre-process the Twitter textual data to ensure that its lexical items are to a high degree lexically comparable to those of natural language. This is done by performing sentiment-aware tokenization, spell correction, word normalization, segmentation (for splitting hashtags) and annotation. The implemented tokenizer with \textit{SentiWordnet} corpus \cite{esuli2006sentiwordnet} is able to avoid splitting expressions or words that should be kept intact (as one token), and identify most emoticons, emojis, expressions such as dates, currencies, acronyms, censored words (e.g. s**t), etc. In addition, we perform modifications on the extracted tokens. For spelling correction, we compose a dictionary for the most commonly seen abbreviations, censored words and elongated words (for emphasis, e.g. "reallyyy"). The \textit{Viterbi} algorithm is used for word segmentation, with word statistics (unigrams and bigrams) computed from the \textit{NLTK English Corpus} to obtain the most probable segmentation posteriors from the unigrams and bigrams probabilities. Moreover, all texts are lower-cased, and URLs, emails and mentioned usernames are replaced with common designated tags so that they would not need to be annotated by the semantic parser.

\subsection{College Metadata}
The  meta-statistics on the college demographics regarding enrollment, geographical location, private/public categorization and male-to-female ratio are obtained. Furthermore, we acquire the Campus Safety and Security Survey dataset from the official U.S. Department of Education website and use rape-related cases statistic as an attribute to complete the data for our linear regression model. The number of such reported cases by these 200 colleges in 2015 amounts to 2,939.

\section{Methodology}
\subsection{Regression Analysis}
We examine other features regarding the characteristics of the studied colleges, which might be significant factors of sexual harassment. Four factual attributes pertaining to the 200 colleges are extracted from the U.S. News Statistics, which consists of Undergraduate  Enrollment, Male/Female Ratio, Private/Public, and Region (Northeast, South, West, and Midwest). We also use the normalized rape-related cases count (number of cases reported per student enrolled) from the stated government resource as another attribute to examine the proximity of our dataset to the official one. This feature vector is then fitted in a linear regression to predict the normalized \#metoo users count (number of unique users who posted \#MeToo tweets per student enrolled) for each individual college.

\subsection {Labeling Sexual Harassment}
Per our topic modeling results, we decide to look deeper into the narratives of \#MeToo users who reveal their personal stories. We examine 6,760 tweets from the most relevant topic of our LDA model, and categorize them based on the following metrics: harassment types (verbal, physical, and visual abuse) and context (peer-to-peer, school employee or work employer, and third-parties). These labels are based on definitions by the U.S. Dept. of Education \cite{home_2020}.

\subsection{Topic Modeling on \#MeToo Tweets}
In order to understand the latent topics of those \#MeToo tweets for college followers, we first utilize Latent Dirichlet Allocation (LDA) to label universal topics demonstrated by the users. We determine the optimal topic number by selecting the one with the highest coherence score. Since certain words frequently appear in those \#MeToo tweets (e.g., sexual harassment, men, women, story, etc.), we transform our corpus using \textit{TF-IDF}, a term-weighting scheme that discounts the influence of common terms.

\subsection{Semantic Parsing with \textit{TRIPS}}
Learning deep meaning representations, which enables the preservation of rich semantic content of entities, meaning ambiguity resolution and partial relational understanding of texts, is one of the challenges that the \textit{TRIPS} parser \cite{allen2018putting} is tasked to tackle. This kind of meaning is represented by \textit{TRIPS} \textit{Logical Form (LF)}, which is a graph-based representation that serves as the interface between structural analysis of text (i.e., parse) and the subsequent use of the information to produce knowledge. The \textit{LF} graphs are obtained by using the semantic types, roles and rule-based relations defined by the \textit{TRIPS Ontology} \cite{allen2018putting} at its core in combination with various linguistic techniques such as Dialogue Act Identification, Dependency Parsing, Named Entity Recognition, and Crowd-sourced Lexicon (Wordnet).

\begin{figure}[h]
\vspace{-6pt}
\centering
\includegraphics[width=8.5cm, height=5.5cm]{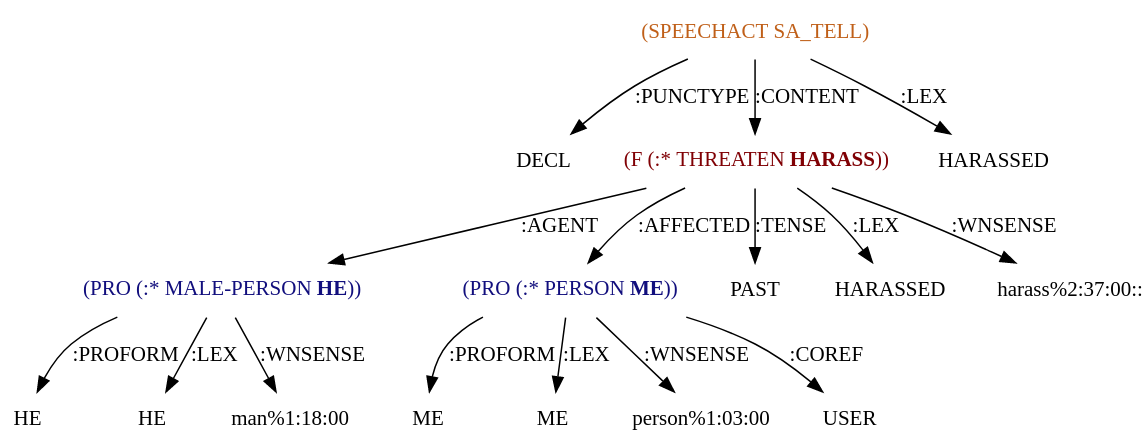}
\vspace{-8pt}
\caption{The meaning representation of the example sentence "He harassed me." in \textit{TRIPS LF}, the \textit{Ontology} types of the words are indicated by ":*" and the role-argument relations between them are denoted by named arcs.}
\centering
\vspace{-0pt}
\end{figure}

Figure 1 illustrates an example of the \textit{TRIPS LF} graph depicting the meaning of the sentence "He harassed me," where the event described though the speech act \textbf{TELL} (i.e. telling a story) is the verb predicate \textbf{HARASS}, which is caused by the agent \textbf{HE} and influences the affected (also called "theme" in traditional literature) \textbf{ME}. As seen from the  previously discussed example, the action-agent-affected relational structure is applicable to even the simplest sentences used for storytelling, and it is in fact very common for humans to encounter in both spoken and written languages. This makes it well suited for event extraction from short texts, useful for analyzing tweets with Twitter's 280 character limit. Therefore, our implementation of \textit{TRIPS} parser is particularly tailored for identifying the verb predicates in tweets and their corresponding agent-affected arguments (with $82.4\%$ \textit{F1} score), so that we can have a solid ground for further analysis. 

\subsection{Connotation Frames and Sentiment Analysis}
In order to develop an interpretable analysis that focuses on sentiment scores pertaining to the entities and events mentioned in the narratives, as well as the perceptions of readers on such events, we draw from existing literature on connotation frames: a set of verbs annotated  according to what they imply about  semantically dependent entities. Connotation frames, first introduced by Rashkin, Singh, and Choi (2016), provides a framework for analyzing nuanced dimensions in text by combining polarity annotations with frame semantics (Fillmore 1982). More specifically, verbs are annotated across various dimensions and perspectives so that a verb might elicit a positive sentiment for its subject (i.e. sympathy) but imply a negative effect for its object. We target the sentiments towards the entities and verb predicates through a pre-collected set of 950 verbs that have been annotated for these traits, which can be more clearly demonstrated through the example "He harassed me.":
\begin{itemize}
    \item ${Sentiment(\textrm{verb}) -}$: something negative happened to the writer.
    \item $Sentiment(\textrm{affected}) -$: the writer (affected) most likely feels negative about the event.
    \item $Perspective(\textrm{affected} \rightarrow \textrm{agent})-$: the writer most likely has negative feelings towards the agent as a result of the event.
    \item $Perspective(\textrm{reader} \rightarrow \textrm{affected})-$: the reader most likely view the agent as the antagonist.
    \item $Perspective(\textrm{affected} \rightarrow \textrm{affected})+$: the reader most likely feels sympathetic towards the writer.
\end{itemize}

In addition to extracting sentiment scores from the pre-annotated corpus, we also need to predict sentiment scores of unknown verbs. To achieve this task, we rely on the 200-dimensional \textit{GloVe} word embeddings \cite{pennington2014glove}, pretrained on their Twitter dataset, to compute the scores of the nearest neighboring synonyms contained in the annotated verb set and normalize their weighted sum to get the resulting sentiment (Equation 1).
\begin{equation} 
\label{eqn:hello}
S(w) =  \mathcal{I}\mathcal{S}(w)+(1-\mathcal{I})\frac{1}{n}\sum_{e\in \gamma(w)\cap \mathcal{A}} Pr(w\rightarrow e)\mathcal{S}(e)
\end{equation}
where $\mathcal{I}=\mathbf{1_{w \in \mathcal{A}}}$ is the indicator function for whether verb predicate $w$ is in the annotation set $\mathcal{A}$, $\gamma(w)$ is the set of nearest neighbors $e$'s of verb $w$. Because our predictive model computes event-entity sentiment scores and generates verb predicate knowledge simultaneously, it is sensitive to data initialization. Therefore, we train the model iteratively on a number of random initialization to achieve the best results.

\section{Experimental Results}
\subsection{Topical Themes of \#MeToo Tweets}

The results of LDA on \#MeToo tweets of college users (Table 1) fall into the same pattern as the research of Modrek and  Chakalov (2019), which suggests that a large portion of \#MeToo tweets on Twitter focuses on sharing personal traumatic stories about sexual harassment \cite{modrek2019metoo}. In fact, in our top 5 topics, Topics 1 and 5 mainly depict gruesome stories and childhood or college time experience. This finding seems to support the validity of the Twitter sample of Modrek and Chakalov (2019), where 11\% discloses personal sexual harassment memories and 5.8\% of them was in formative years~\cite{modrek2019metoo}. These users also shows multiple emotions toward this movement, such as compassion (topic 2), determination (topic 3), and hope (topic 4). We will further examine the emotion features in the latter results.

\begin{table}
\begin{tabular}{|p{0.6cm}|p{1.9cm}|p{5.2cm}|}
\hline
Topic & Topic Label & Keywords \\
\hline\hline
1 & Recalling the experience & times, old, man, got, called, girl, wrong, asked, home, done, scared, tried,  room\\
\hline

2 & Showing sympathy, sharing news & sexually, harassed, assaulted, women, experiences, magnitude, sense, problem, fear, men, status, getting, heartbreaking\\
\hline

3 & Calling for actions & story, normal, denial, please, forward, tell, shame, shared, wish, women, daughters, worse, glad\\
\hline
4 & Showing optimism & instant, take, right, fight, past, action, self, write, hey, loved, spoke, sisters, body, former, front, claims, stronger\\
\hline

5 & Detailing early life experience & call, issue, hashtag, child, awareness, guy, telling, trauma, number, party, teacher, sexual, raise, sometimes\\
\hline

\end{tabular}
\smallskip
\caption{Top 5 topics from all \#MeToo Tweets from 51,104 college followers.}
\vspace*{-2mm}
\end{table}

\subsection{Regression Result}
\begin{table}[h!]
\centering
\begin{tabular}{|p{1.9cm}| c| c| c| c|} 
\hline
Feature & Coefficient & Std. Err. & t-stat & p-value \\ [0.5ex] 
\hline\hline
M/F Ratio & 3.613e-04 & 6.690e-03 & 0.054 & 0.9570 \\
\hline
Enrollment & 1.075e-06 & 1.639e-06 & 0.656 & 0.5128 \\
\hline
Private & 2.858e-03 & 3.676e-02 & 0.078 & 0.9381 \\
\hline
Northeast & \textbf{7.840e-02} & 3.734e-02 & 2.100 & \textbf{0.0370} \\
\hline
West & \textbf{8.909e-02} & 4.032e-02 & 2.210 & \textbf{0.0283} \\
\hline
South & \textbf{7.529e-02} & 3.763e-02 & 2.001 & \textbf{0.0468} \\
\hline
Normalized cases count & \textbf{9.098e+01} & 1.176e+01 & 7.735 & \textbf{5.7e-13} \\
\hline
constant & -7.943e-02 & 4.990e-02 & -1.592 & 0.1131 \\
\hline
\end{tabular}
\caption{Linear regression results.} 
\end{table}

Observing the results of the linear regression in Table 2, we find the normalized governmental reported cases count and regional feature to be statistically significant on the sexual harassment rate in the Twitter data ($p-value<0.05$). Specifically, the change in the number of reported cases constitutes a considerable change in the number of \#MeToo users on Twitter as \textit{p-value} is extremely small at $5.7e-13$. This corresponds to the research by Napolitano (2014) regarding the "Yes means yes" movement in higher education institutes in recent years, as even with some limitations and inconsistency, the sexual assault reporting system is gradually becoming more rigorous \cite{napolitano2014only}.  Meanwhile, attending colleges in the Northeast, West and South regions increases the possibility of posting about sexual harassment (positive coefficients), over the Midwest region. This finding is interesting and warrants further scrutiny.

\subsection{Event-Entity Sentiment Analysis}

We discover that approximately half of users who detailed their sexual harassment experiences with the \#MeToo hashtag suffered from physical aggression. Also, more than half of them claimed to encounter the perpetrators outside the college and work environment.  The sentimental score for the affected entities and the verb of cases pertaining to faculty are strictly negative, suggesting that academic personnel's actions might be described as more damaging to the students' mental health. This finding resonates a recent research by Cantapulo et al. regarding the potential hazard of sexual harassment conducts by university faculties using data from federal investigation and relevant social science literature~\cite{cantalupo2018systematic}. Furthermore, many in this group tend to mention their respective age, typically between 5 and 20 (24\% of the studied subset). This observation reveals an alarming number of child and teenager sexual abuse, indicating that although college students are not as prone to sexual harassment from their peers and teachers, they might still be traumatized by their childhood experiences.

In addition, although verbal abuse experiences accounts for a large proportion of the tweets, it is challenging to gain sentiment insights into them, as the majority of them contains insinuations and sarcasms regarding sexual harassment. This explains why the sentiment scores of the events and entities are very close to neutral.

\begin{table}[h!]
\centering
\begin{tabular}{|p{1.5cm}|p{1.4cm}|p{1.3cm}| p{1.3cm}| p{1.3cm}| p{1.3cm}|} 
\hline
Harassment  &  Participant &	Event-Sentiment	& Affected-Sentiment & Percentage\\
\hline\hline
Physical	& 3rd-Party & -0.0429	& -0.0479 & 23.63\%\\
\hline
Physical	& Faculty & -0.1999	& -0.2308 & 11.39\%\\
\hline
Physical	& Peer & -0.0136 & -0.1018 & 16.03\%\\
\hline
Verbal	& 3rd-Party & 0.1385 & 0.0077	& 25.32\%\\
\hline
Verbal	& Faculty & 0.1454 & 0.0051	& 5.91\%\\
\hline
Verbal	& Peer & 0.0819	& -0.0024 & 5.91\%\\
\hline
Visual	& 3rd-Party & 0.1015 & -0.0407 & 6.33\%\\
\hline
Visual	& Faculty & 0.1333 & 0.0500	& 0.42\%\\
\hline
Visual	& Peer & -0.3946 & 0.0000 & 0.84\%\\
\hline
\end{tabular}
\caption{Semantic sentiment results.} 
\end{table}

\subsection{Limitations and Ethical Implications}

Our dataset is taken from only a sample of a specific set of colleges, and different samples might yield different results. Our method of identifying college students is simple, and might not reflect the whole student population. Furthermore, the majority of posts on Twitter are short texts (under 50 words). This factor, according to previous research, might hamper the performance of the \textit{LDA} results, despite the use of the \textit{TF-IDF} scheme \cite{tang2014understanding}.

Furthermore, while the main goal of this paper is to shed lights to the ongoing problems in the academia and contribute to the future sociological study using big data analysis, our dataset might be misused for detrimental purposes. Also, data regarding sexual harassment is sensitive in nature, and might have unanticipated effects on those addressed users.

\section{Conclusion}

In this study, we discover a novel correlation between the number of college users who participate in the \#MeToo movement and the number of official reported cases from the government data. This is a positive sign suggesting that the higher education system is moving into a right direction to effectively utilize Title IV, a portion of the Education Amendments Act of 1972,~\footnote{https://goo.gl/J5ZSpb} which requests colleges to submit their sexual misconduct reports to the officials and protect the victims.   In addition, we capture several geographic and behavioral characteristics of the \#MeToo users related to sexual assault such as region, reaction and narrative content following the trend, as well as sentiment and social interactions, some of which are supported by various literature on sexual harassment. Importantly, our semantic analysis reveals interesting patterns of the assaulting cases. We believe our methodologies on defining these \#MeToo users and their features will be applicable to further studies on this and other alarming social issues.

Furthermore, we find that the social media-driven approach is highly useful in facilitating crime-related sociology research on a large scale and spectrum. Moreover, since social networks appeal to a broad audience, especially those outside academia, studies using these resources are highly useful for raising awareness in the community on concurrent social problems.

Last but not least, many other aspects of the text data from social media, which could provide many interesting insights on sexual harassment, remain largely untouched. In the future, we intend to explore more sophisticated language features and implement more supervised models with advanced neural network parsing and classification. We believe that with our current dataset, an extension to take advantage of cutting-edge linguistic techniques will be the next step to address the previously unanswered questions and uncover deeper meanings of the tweets on sexual harassment.

\bibliography{aaai}

\begin{thebibliography}{}

\bibitem[\protect\citeauthoryear{Allen and Teng}{2018}]{allen2018putting}
Allen, J., and Teng, C.~M.
\newblock 2018.
\newblock Putting semantics into semantic roles.
\newblock In {\em Proceedings of the Seventh Joint Conference on Lexical and
  Computational Semantics},  235--244.

\bibitem[\protect\citeauthoryear{Blandfort \bgroup et al\mbox.\egroup
  }{2019}]{blandfort2019multimodal}
Blandfort, P.; Patton, D.~U.; Frey, W.~R.; Karaman, S.; Bhargava, S.; Lee,
  F.-T.; Varia, S.; Kedzie, C.; Gaskell, M.~B.; Schifanella, R.; et~al.
\newblock 2019.
\newblock Multimodal social media analysis for gang violence prevention.
\newblock In {\em Proceedings of the International AAAI Conference on Web and
  Social Media}, volume~13,  114--124.

\bibitem[\protect\citeauthoryear{Brutus, Aguinis, and
  Wassmer}{2013}]{doi:10.1177/0149206312455245}
Brutus, S.; Aguinis, H.; and Wassmer, U.
\newblock 2013.
\newblock Self-reported limitations and future directions in scholarly reports:
  Analysis and recommendations.
\newblock {\em Journal of Management} 39(1):48--75.

\bibitem[\protect\citeauthoryear{Cantalupo and
  Kidder}{2018}]{cantalupo2018systematic}
Cantalupo, N.~C., and Kidder, W.~C.
\newblock 2018.
\newblock A systematic look at a serial problem: Sexual harassment of students
  by university faculty.
\newblock {\em Utah L. Rev.}  671.

\bibitem[\protect\citeauthoryear{Cantu}{2020}]{home_2020}
Cantu, N.
\newblock 2020.
\newblock Sexual harassment guidance.

\bibitem[\protect\citeauthoryear{ElSherief \bgroup et al\mbox.\egroup
  }{2018}]{elsherief2018hate}
ElSherief, M.; Kulkarni, V.; Nguyen, D.; Wang, W.~Y.; and Belding, E.
\newblock 2018.
\newblock Hate lingo: A target-based linguistic analysis of hate speech in
  social media.
\newblock In {\em Twelfth International AAAI Conference on Web and Social
  Media}.

\bibitem[\protect\citeauthoryear{Esuli and
  Sebastiani}{2006}]{esuli2006sentiwordnet}
Esuli, A., and Sebastiani, F.
\newblock 2006.
\newblock Sentiwordnet: A publicly available lexical resource for opinion
  mining.
\newblock In {\em LREC}, volume~6,  417--422.
\newblock Citeseer.

\bibitem[\protect\citeauthoryear{Field, Bhat, and
  Tsvetkov}{2019}]{field2019contextual}
Field, A.; Bhat, G.; and Tsvetkov, Y.
\newblock 2019.
\newblock Contextual affective analysis: A case study of people portrayals in
  online\# metoo stories.
\newblock In {\em Proceedings of the International AAAI Conference on Web and
  Social Media}, volume~13,  158--169.

\bibitem[\protect\citeauthoryear{Fitzgerald \bgroup et al\mbox.\egroup
  }{1988}]{FITZGERALD1988152}
Fitzgerald, L.~F.; Shullman, S.~L.; Bailey, N.; Richards, M.; Swecker, J.;
  Gold, Y.; Ormerod, M.; and Weitzman, L.
\newblock 1988.
\newblock The incidence and dimensions of sexual harassment in academia and the
  workplace.
\newblock {\em Journal of Vocational Behavior} 32(2):152 -- 175.

\bibitem[\protect\citeauthoryear{Fitzgerald, Gelfand, and
  Drasgow}{1995}]{doi:10.1207/s15324834basp17042}
Fitzgerald, L.~F.; Gelfand, M.~J.; and Drasgow, F.
\newblock 1995.
\newblock Measuring sexual harassment: Theoretical and psychometric advances.
\newblock {\em Basic and Applied Social Psychology} 17(4):425--445.

\bibitem[\protect\citeauthoryear{He, Murphy, and
  Luo}{2016}]{DBLP:journals/corr/HeML16}
He, L.; Murphy, L.; and Luo, J.
\newblock 2016.
\newblock Using social media to promote {STEM} education: Matching college
  students with role models.
\newblock {\em CoRR} abs/1607.00405.

\bibitem[\protect\citeauthoryear{Huerta \bgroup et al\mbox.\egroup
  }{2006}]{doi:10.1177/0146167205284281}
Huerta, M.; Cortina, L.~M.; Pang, J.~S.; Torges, C.~M.; and Magley, V.~J.
\newblock 2006.
\newblock Sex and power in the academy: Modeling sexual harassment in the lives
  of college women.
\newblock {\em Personality and Social Psychology Bulletin} 32(5):616--628.
\newblock PMID: 16702155.

\bibitem[\protect\citeauthoryear{Magu, Joshi, and
  Luo}{2017}]{DBLP:journals/corr/MaguJL17}
Magu, R.; Joshi, K.; and Luo, J.
\newblock 2017.
\newblock Detecting the hate code on social media.
\newblock {\em CoRR} abs/1703.05443.

\bibitem[\protect\citeauthoryear{McDermut, Haaga, and Kirk}{2000}]{JTS:JTS213}
McDermut, J.~F.; Haaga, D. A.~F.; and Kirk, L.
\newblock 2000.
\newblock An evaluation of stress symptoms associated with academic sexual
  harassment.
\newblock {\em Journal of Traumatic Stress} 13(3):397--411.

\bibitem[\protect\citeauthoryear{Modrek and Chakalov}{2019}]{modrek2019metoo}
Modrek, S., and Chakalov, B.
\newblock 2019.
\newblock The \#metoo movement in the united states: text analysis of early
  twitter conversations.
\newblock {\em Journal of medical internet research} 21(9):e13837.

\bibitem[\protect\citeauthoryear{Napolitano}{2014}]{napolitano2014only}
Napolitano, J.
\newblock 2014.
\newblock Only yes means yes: An essay on university policies regarding sexual
  violence and sexual assult.
\newblock {\em Yale L. \& Pol'y Rev.} 33:387.

\bibitem[\protect\citeauthoryear{Pennington, Socher, and
  Manning}{2014}]{pennington2014glove}
Pennington, J.; Socher, R.; and Manning, C.
\newblock 2014.
\newblock Glove: Global vectors for word representation.
\newblock In {\em Proceedings of the 2014 conference on empirical methods in
  natural language processing (EMNLP)},  1532--1543.

\bibitem[\protect\citeauthoryear{Tang \bgroup et al\mbox.\egroup
  }{2014}]{tang2014understanding}
Tang, J.; Meng, Z.; Nguyen, X.; Mei, Q.; and Zhang, M.
\newblock 2014.
\newblock Understanding the limiting factors of topic modeling via posterior
  contraction analysis.
\newblock In {\em International Conference on Machine Learning},  190--198.

\bibitem[\protect\citeauthoryear{Timothy \bgroup et al\mbox.\egroup
  }{1982}]{doi:10.1111/j.1540-4560.1982.tb01912.x}
Timothy, R.; Sandra, C.; Valerie, D.; Kim, B.; and B., B.~M.
\newblock 1982.
\newblock The factorial survey: An approach to defining sexual harassment on
  campus.
\newblock {\em Journal of Social Issues} 38(4):99--110.

\bibitem[\protect\citeauthoryear{Wang \bgroup et al\mbox.\egroup
  }{2016}]{DBLP:journals/corr/WangFZL16}
Wang, Y.; Feng, Y.; Zhang, X.; and Luo, J.
\newblock 2016.
\newblock Gender politics in the 2016 {U.S.} presidential election: {A}
  computer vision approach.
\newblock {\em CoRR} abs/1611.02806.

\end{thebibliography}
\bibliographystyle{aaai}
\end{document}